\documentclass[twocolumn,prl,preprintnumbers,nofootinbib]{revtex4}
\usepackage[utf8]{inputenc}
\usepackage{amsmath}
\usepackage{amssymb}
\usepackage{graphicx}
\usepackage{physics}
\usepackage{comment}
\usepackage{booktabs}
\usepackage{braket}
\usepackage{bm}
\usepackage{bbm}
\usepackage{epsfig}
\usepackage{pstricks}
\usepackage{color}
\usepackage{tikz}
\usepackage{subfig}
\usepackage{cancel}

\newcommand{\be}{\begin{eqnarray}}
\newcommand{\ee}{\end{eqnarray}}

\newcommand{\aln}[1]{\begin{align}#1\end{align}}
\newcommand{\gat}[1]{\begin{gather}#1\end{gather}}

\begin{document}

\preprint{KEK-TH-2604}

\title{Towards a spatial cat state of a massive pendulum} 
\author{Satoshi Iso$^{1,2,3,4 )}$ }
\author{Jinyang Li $^{1,3)}$}
\author{  Nobuyuki Matsumoto$^{5)}$ } 
\author{ Katsuta Sakai $^{1,6)}$}
\affiliation{
$^1$ KEK Theory Center,  Institute of Particle and Nuclear Studies  \\
$^2$ International Center for Quantum-field Measurement Systems for Studies of the Universe and Particles,   
 High Energy Accelerator Research Organization (KEK), 
  1-1 Oho, Tsukuba, Ibaraki 305-0801, Japan \\
$^3$ Graduate University for Advanced Studies (SOKENDAI), Oho 1-1, Tsukuba, Ibaraki 305-0801, Japan. \\
$^4$ RIKEN Interdisciplinary Theoretical and Mathematical Sciences (iTHEMS), Wako, Saitama 351-0198, Japan \\
$^5$ Department of Physics, Gakushuin University, 1-5-1 Mejiro, Toshima, Tokyo 171-8588, Japan \\
$^6$ College of Liberal Arts and Sciences, Tokyo Medical and Dental University,  2-8-30 Kounodai, Ichikawa, Chiba 272-0827, Japan
}

\begin{abstract} 

We propose an experiment for constructing a spatial cat state of a suspended mirror with an order of $\cal{O}$(mg). 
The mirror is set at the center of two mirrors, creating two optical cavities and optical springs. 
 The induced potential exhibits a double-well shape, and its deformation resembles a second-order phase transition as a function of laser power. 
 We estimate an adiabatic condition for the ground state wave function to metamorphose from a localized state at the origin to a spatial cat state within the double-well potential, within a coherence time determined by mechanical and environmental noises. Our  estimation suggests that such a construction is possible if we can provide an ultra-high finesse optical cavity with $\mathcal{F}=2.5\times10^5$ and a length of $0.3$ cm, along with a shot-noise-limited laser at 7.9 nW. The necessary mechanical coherence time is approximately one second.

\end{abstract}

\maketitle


{\it Introduction}.---
A macroscopic quantum state has long been attracting a lot of attention since Schrödinger's Gedanken-experiment \cite{Schroedinger1935}. 
In particular, a spatial cat state, which is a superposition of spatially distant states, has been widely studied in systems including an atomic fountain \cite{PhysRevLett.63.612}, electrons \cite{tonomura1989}, molecules \cite{Fein2019}, a superconducting device \cite{friedman2000quantum}, and an optomechanical mirror \cite{Marshall:2002exi}.
Recently, a spatial cat of a massive object attracts much attention since it would be helpful to reveal quantum nature of gravity \cite{PhysRevLett.119.240401,PhysRevLett.119.240402}.  
In general relativity, gravity is understood via spacetime. A key assumption in quantum gravity is the existence of a quantum superposition of spacetime, crucial for understanding the fundamental property of spacetime. 
However, there has been no demonstration of a spatial cat state of a massive object, leaving verification of this aspect elusive.


In this letter, we propose a novel method for constructing a spatial cat state of a massive object, such as a 1 mg mirror, based on optomechanical coupling with cavity photons. This coupling induces an effective potential of the mirror, known as the optical spring effect. Typically, linear analysis regarding the mirror's displacement is employed and widely utilized for cooling purposes. On the other hand, nonlinear analysis is shown to yield a potential with multiple minima depending on the laser power \cite{Marquardt_2006}. This presents the possibility of constructing a cat state of the mirror. However, achieving this requires further steps. Specifically, we need to be able to adiabatically transform the potential from a simple harmonic type with a single minimum to a double-well shape.  
Furthermore, this deformation must resemble a second-order phase transition concerning the laser power. In other words, a potential barrier should be absent when nontrivial minima emerge at specific magnitudes of laser power.
To construct a cat state, a ground state of the harmonic potential is prepared at the origin and then gradually increase the laser power. By slowly deforming the potential into a double-well type within the coherence time, the ground state localized at the origin will metamorphose into a cat state, representing the ground state of the new potential with two minima.

\begin{figure}[ht]
\centering
\includegraphics[width=0.75\hsize]{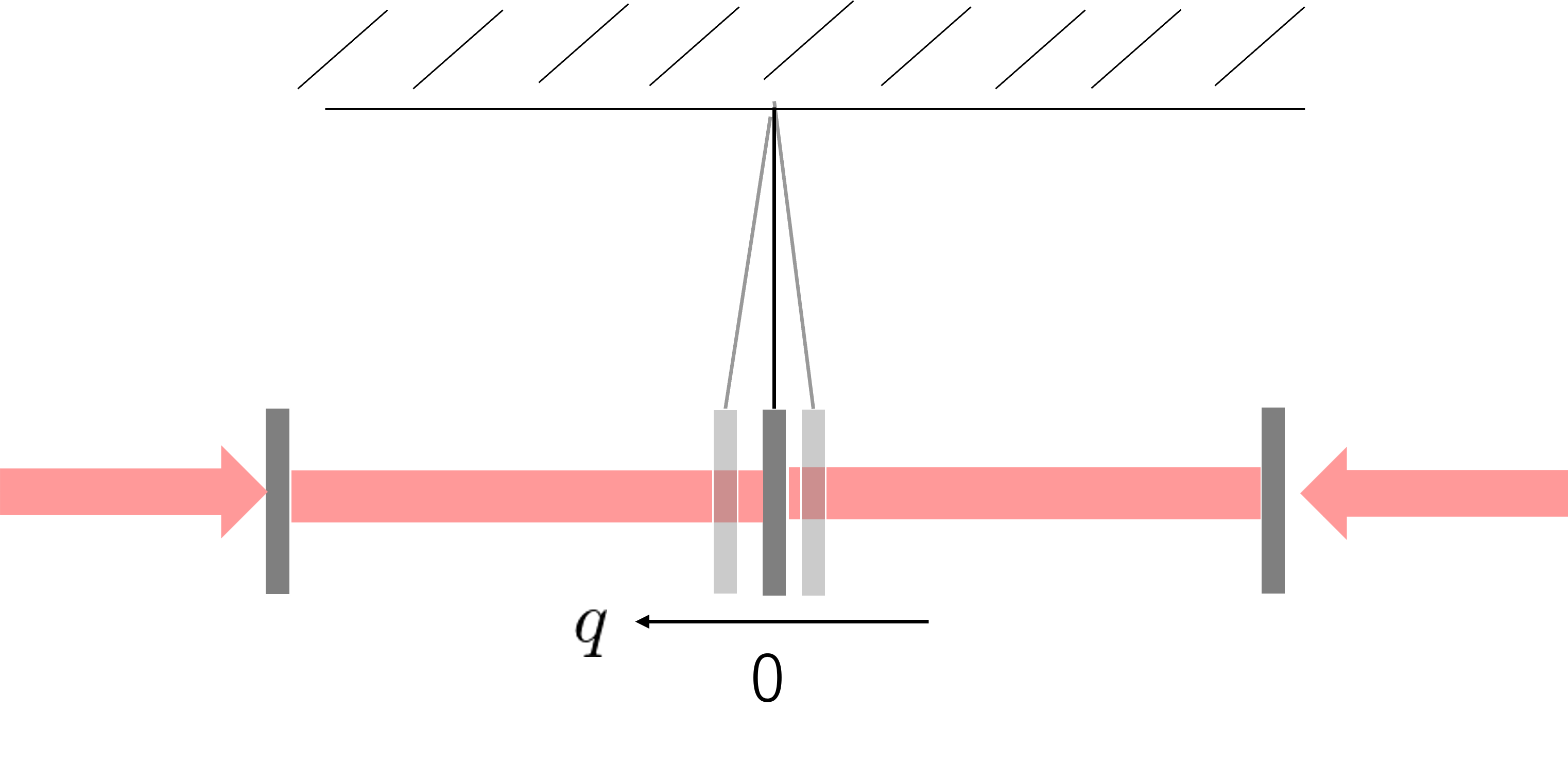}
\centering
\caption{\footnotesize  
An optomechanical system with dual lasers. All parameters on the left and right sides are tuned to be identical.}
\label{fig: optomecha_double}
\end{figure}

\vspace{1mm}
{\it  Experimental setup and Hamiltonian}.---  
Our experimental setup is illustrated in Fig. \ref{fig: optomecha_double}. We introduce two lasers from both sides of a suspended mirror positioned in the center of a cavity.
The Hamiltonian is given by 
\aln{
\frac{H}{\hbar} = & \frac{\Omega}{2}(\hat{p}^2+\hat{q}^2) + \sum_{i=L,R} \omega_c \hat{a_i}^\dag \hat{a_i} \nonumber \\
 & +i F \sum_{i=L,R}  (e^{-i\omega_lt}\hat{a_i}^\dag-e^{i\omega_lt}\hat{a_i})\nonumber\\
 & +g \hat{q}(\hat{a_L}^\dag \hat{a_L}-\hat{a_R}^\dag \hat{a_R}),
 \label{eq: Hamiltonian}
}
where $(a_{L},a_{L}^\dag)$ and $(a_{R},a_{R}^\dag)$ are the creation and annihilation operators of the optical cavity modes on the left and right, respectively, while $(q,,p)$ represent the dimensionless position and momentum operators of the mechanical modes. To ensure a symmetric potential, various parameters are adjusted to be identical between the left and right sides. The size of the ground state wave function of the mechanical object, mirror, is given by $\langle q^2 \rangle = \langle p^2 \rangle=1/2$. $\Omega$ denotes the frequency of the mechanical mode, and $\omega_c$ is the resonant frequency of the cavity of length $L$. The third term represents the driving force exerted by an external laser of frequency $\omega_l$, where $F$ stands for the effective amplitude of the driving laser.

The fourth term arises from the change in the resonant frequency of the cavity due to the displacement of the mirror. Here, $g$ denotes the optomechanical coupling constant, defined as $g=G\sqrt{\hbar/m\Omega}$, with $G=\omega_c/L :=(\omega_l +\Delta)/L$ representing the optical frequency shift per unit displacement. $\Delta$ represents the detuning frequency, and $m$ denotes the mechanical mass.

Additionally, dissipative loss rates for the cavity and mechanical modes are introduced, denoted by $\kappa$ and $\gamma_m$, respectively. $\kappa$ is the sum of an internal loss rate $\kappa_i$ and an external one $\kappa_e$. These rates can be incorporated into the master equation for density matrix evolution.

\vspace{1mm}
{\it  Effective potential}.--- 
Radiation pressure generates an effective potential for the mirror. To derive this potential, we assume a hierarchy in the frequencies of the mechanical mode $\Omega$ and $\gamma_m$, and the cavity modes $\Delta$ and $\kappa$ as follows:
\aln{
\Omega,\,\gamma_m\ll \Delta,\,\kappa  
}
and adopt the Born-Oppenheimer approximation. 
By integrating out the cavity modes, the effective potential of the mirror is given by \cite{Marquardt_2006}:
\aln{
\frac{V(q; n)}{\hbar}&\simeq\frac{\Omega}{2}q^2+n\frac{\Delta^2+\kappa^2/4}{\kappa/2}
\sum_{\epsilon=\pm} \mathrm{Arctan}\left(\frac{\Delta+ \epsilon gq}{\kappa/2}\right) \ .
\nonumber
}
Here $\epsilon =\pm$ corresponds to contributions from the left and the right lasers, respectively. $n$ represents the number of photons in each of cavities, defined as:
\aln{
n=\frac{F^2}{\Delta^2+\kappa^2/4}.
\label{eq: determination of n}
}
\begin{figure}[t]
\centering
\begin{minipage}[t]{0.5\textwidth}
\includegraphics[width=0.9\hsize]{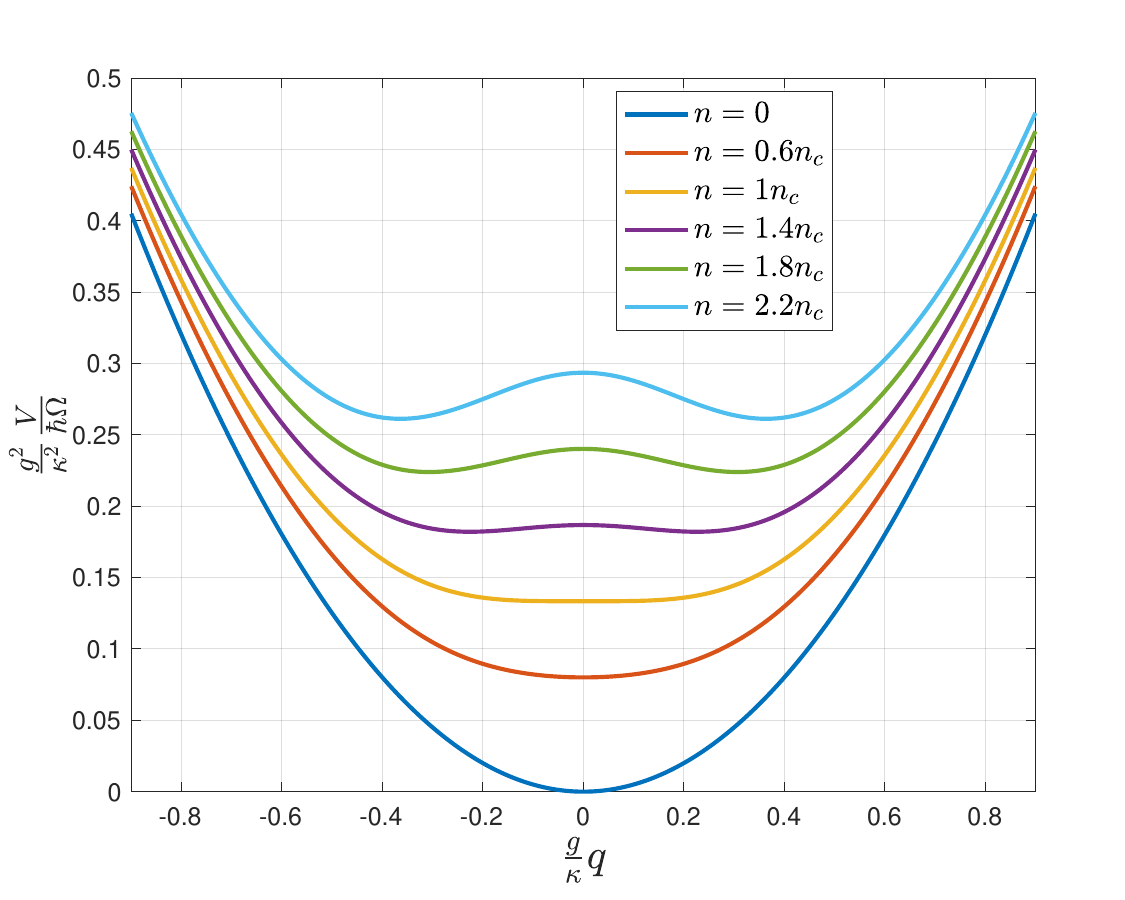}  
\end{minipage}
\centering
\caption{\footnotesize 
Varying the laser power $n$ deforms the potential $V(q;n)$. 
Here, we choose $\Delta = 0.1 \kappa$, satisfying the condition \eqref{Eq:D<k}. This deformation resembles a second-order phase transition. The six different curves represent various values of $n$, respectively.
}
\label{fig: double2}
\end{figure}
While we have written the driving laser as a classic field in Eq.\,\eqref{eq: Hamiltonian}, we will later take into account the vacuum fluctuation coming from the laser port, which results in a Poissonian fluctuation of $n$.
Additionally, the friction term of the mechanical mode $\gamma_m$ is modified and depends on the position $q$ of the mirror. For small displacements $g|q| \ll |\Delta|$, it is approximately given by a position-independent constant shift as follows:
\aln{
\Gamma 
&=\gamma_m+  n \frac{ 4g^2 \Omega\kappa \Delta}{[\Delta^2+\kappa^2/4]^2} \ .
\label{eq: Gamma_c_single}
}
The second term represents the decoherence rate due to quantum back-action caused by the laser. In the present discussion, as long as $\Omega$ is not very large, this rate is smaller than $\gamma_m$ owing to the hierarchy $(\Omega, g \ll \Delta,\kappa)$.

For a small number of cavity photons, particularly at $n=0$, there is only a single minimum of the effective potential $V(q;n)$ at the origin $q=0$. As $n$ increases, the potential $V(q;n)$ deforms and can have at most three minima.
But, if the condition
\aln{
\Delta \le \frac{\kappa}{2} \hspace{5mm}   \hspace{5mm} \text{(2nd order phase transition)}
\label{Eq:D<k}
}
is satisfied, only two minima exists. 
Namely, a minimum at the origin separates into two minima as shown in Fig. \ref{fig: double2}. We refer to this as the second-order phase transition condition. If this condition is met, we can construct a cat state starting from a localized ground state. 

The critical number of photons in each cavity is given by
\aln{
n_c =  \frac{(4 \Delta^2  + \kappa^2 ) \ \Omega}{16 g^2  \Delta } \ 
\label{Eq:n_c}
}
at which a minimum at the origin separates into two at $q=\pm q_{\text{min}}$. The potential is very flat at the critical point. For $q_{\text{min}} \ll \kappa/g$, the position of the new minima $\pm q_{\text{min}}$ is related to the corresponding number of photons $n(q_{\text{min}})$ as
\aln{
\frac{\delta n} {n_c} :=
\frac{n(q_{\text min}) - n_c}{n_c} &\sim 
\frac{8  g^2 \left(\kappa^2-4 \Delta^2 
\right)}{\left(4 \Delta^2+\kappa^2\right)^2}  q_{\text min}^2 \ .
\label{eq:nqrelation}
}
This relation tells us how much we need to change $n$ from $n_c$ to realize a cat state at $q=\pm q_{\text{min}}$. For $|\Delta| \ll \kappa$, the relation becomes $\delta n/n_c \sim 8 (g/\kappa)^2 q_{\text{min}}^2$. The response of $q_{\text{min}}$ to $\delta n/n_c$ becomes larger for smaller $g/\kappa$.

\vspace{1mm}
{\it  Construction of a cat state}.--- 
In  constructing a cat state, we first prepare a localized state at the origin when $n \lesssim n_c$, and then gradually increase the laser power from $n_c$ to $n(q_{\text{min}})$. If the change in the potential is sufficiently slow, the localized ground state can adiabatically evolve into a superposition of the left and right ground states. However, if the change is too rapid, the deformation of the potential excites the ground state to higher excited states, breaking the phase coherence between the left and right wave functions. Therefore, the process must be sufficiently adiabatic. However, the duration of cat state generation is limited by the coherence time determined by mechanical and environmental noises. 

In the following, we investigate the possibility of such an adiabatic construction of a cat state.
\begin{figure}[t]
\centering
\begin{minipage}[t]{0.3\textwidth}
\includegraphics[width=1.22\hsize]{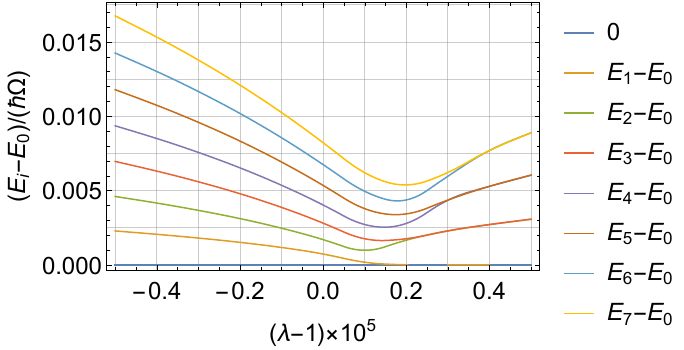} 
\end{minipage}
\begin{minipage}[t]{0.3\textwidth}
\,
\end{minipage}
\begin{minipage}[t]{0.29\textwidth}
\includegraphics[width=1.2\hsize]{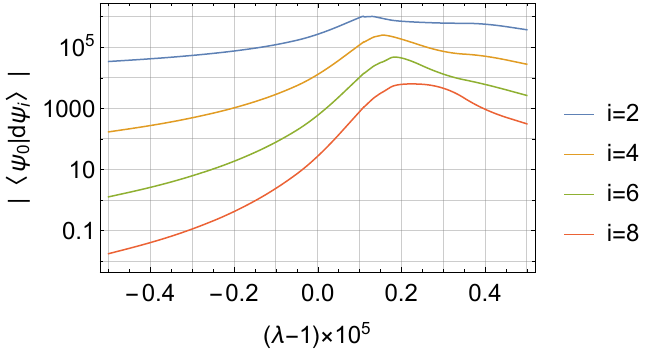}
\end{minipage}
\centering
\caption{\footnotesize Energy eigenvalues of the Hamiltonian (up to 9th excited states) (upper) and overlapping of wave functions of 2,4,6,8th excited states with the ground state (lower) as a function of $\lambda=n/n_c$ at parameters listed in Table \ref{tab: parameters}.} 
\label{fig: adiabatic_mirror}
\end{figure}
Suppose that a Hamiltonian $H(\lambda(t))$ is parametrized by an external parameter $\lambda$, which varies with time $t$. Let the $i$-th eigenvalue and its eigenvectors of $H(\lambda(t))$ be denoted as $E_i(\lambda)$ and $|\psi_i(\lambda) \rangle$, respectively. The typical behavior of the energy eigenvalues of our Hamiltonian is shown in Fig.\ref{fig: adiabatic_mirror}. We are interested in finding the solution of the Schrödinger equation:
$
d|\Psi(t)\rangle /dt =-i H(\lambda(t))/\hbar | \Psi(t) \rangle 
$
with the ground state initial condition. 
Expanding it as $| \Psi(t) \rangle =\sum_i c_i(t) | \psi_i(t) \rangle$, we have the evolution equations for $c_i$;
\aln{
\frac{dc_i}{dt} =-i \frac{E_i(\lambda)}{\hbar} c_i - \sum_j  \dot\lambda \langle  \psi_i|d\psi_j \rangle  c_j 
}
and the initial conditions are given by 
 $c_0(0)=1$ and $c_i(0)=0$ for $i \neq 0$.
Due to the parity symmetry of the potential, only even-parity modes are mixed with the ground state of the harmonic potential at $t=0$. From these equations, we observe that if the off-diagonal terms between $i=0$ and $i \neq 0$ are sufficiently small compared to the difference of diagonal terms $\Delta E_i := E_i-E_0$, we can adiabatically deform the wave function into a cat state. In our case, the parameter is given by $\lambda=n/n_c$.
Thus 
\aln{
\frac{\Delta E_i}{\hbar} > \frac{\dot{n}}{n_c} |\langle \psi_i|d\psi_0 \rangle|
= \frac{\dot{n}}{n_c}  \frac{ |\langle \psi_i|\partial H/\partial \lambda| \psi_0 \rangle |}{\Delta E_i}
\label{eq: nomixing condition}
}
is the necessary condition for adiabatically generating a cat state at each $n$ between $n_c$ and $n(q_{min})$. 
This condition determines the maximum speed at which $n$ can change, ensuring that the ground state wave function can undergo adiabatic transformation into the ground state of the double-well potential. It must be completed before the quantum coherence of the mirror is lost by mechanical and environmental noises.

{\it  Experimental feasibility}.--- 
\begin{table}[b]
  \caption{Typical parameters }
  \label{tab: parameters}
  \centering
  \begin{tabular}{lr}
    Mirror mass & $m = 10$ mg \\
    Mirror frequency & $\Omega/2\pi =1 $\text{ Hz} \\
    Mirror dissipation rate & $ \gamma_m/2\pi = 0.1$ Hz \\ 
    Cavity length & L=  0.06   m\\
    Cavity net loss rate & $\kappa/2\pi = 1 $ MHz \\
    Cavity detuning & $\Delta/2\pi  = 10$ kHz \\
    Frequency shift  & $G/2\pi = \omega_c/L =4.72$ PHz/m \\
    Optomechanical coupling 
   & $ g/2\pi  = 6.10 $ Hz  \\   
    \bottomrule          
  \end{tabular}
\end{table}
\begin{figure}[t]
\begin{tabular}{lcr}
\begin{minipage}[t]{0.2\textwidth}
\centering
\includegraphics[width=1.05\hsize]{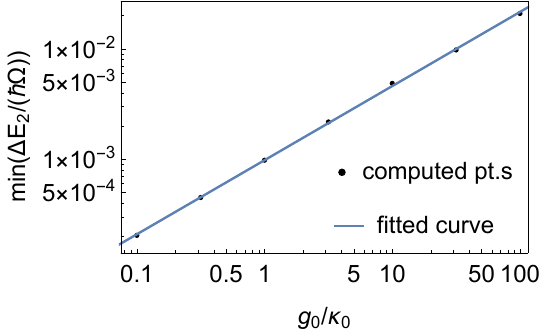}
\end{minipage}
\begin{minipage}[t]{.01\textwidth}
\,
\end{minipage}
\begin{minipage}[t]{.2\textwidth}
\centering
\includegraphics[width=1.05\hsize]{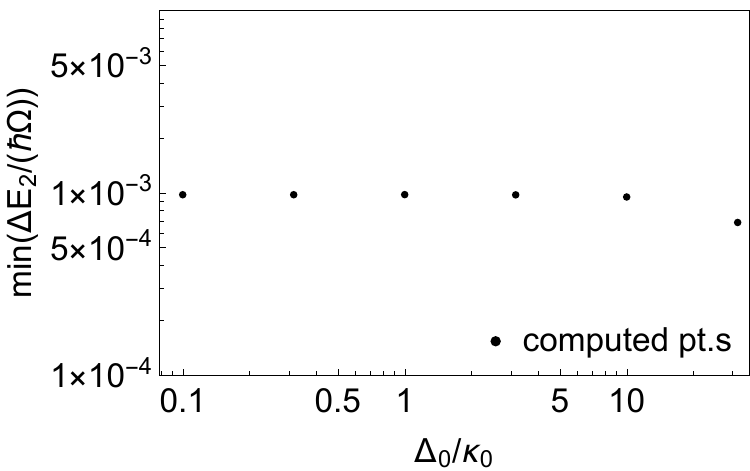}
\end{minipage}\vspace{4pt}\\
\begin{minipage}[t]{0.2\textwidth}
\centering
\includegraphics[width=1.02\hsize]{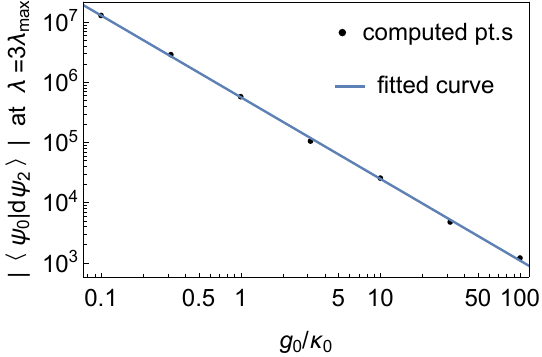}
\end{minipage}
\begin{minipage}[t]{.01\textwidth}
\,
\end{minipage}
\begin{minipage}[t]{.2\textwidth}
\centering
\includegraphics[width=1.09\hsize]{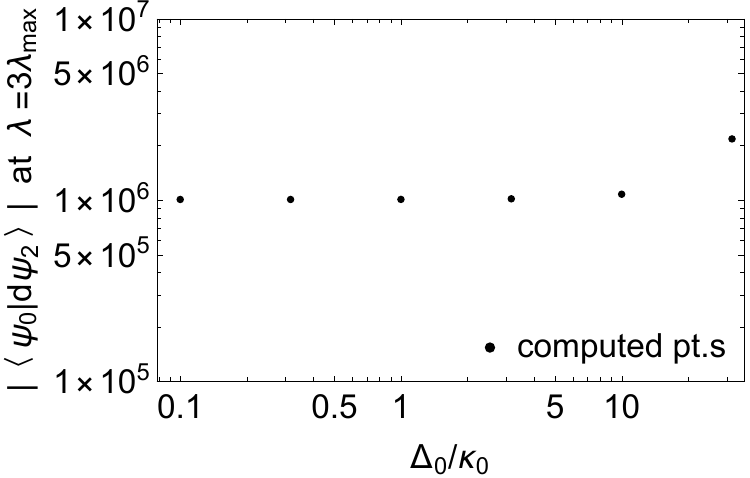}
\end{minipage}
\end{tabular}
\centering
\caption{\footnotesize 
The minimal energy gap $\Delta E_2/(\hbar\Omega)$ (upper) and the asymptotic value of overlap of the wavefunctions $|\langle d_{\lambda} \Psi_2|\Psi_0 \rangle|$ between the ground state and the second excited state (lower) as a function of $g/\kappa$ and $\Delta/\kappa$ are presented. We set $\Delta_0/\kappa_0=1$ for the left figures and $g_0/\kappa_0=1$ for the right ones.
They are well approximated as
$\Delta E/(\hbar\Omega) \sim 9.78\times10^{-4},(g_0/\kappa_0)^{0.672}$
and $|\langle d_{\lambda} \Psi_i|\Psi_0 \rangle |\sim5.65\times10^5,(\kappa_0/g_0)^{1.36}$.
Note that the first excited state has opposite parity to the ground state and remains unmixed.
} 
\label{fig: mixing adiabatic_mirror}
\end{figure}
We will now investigate the experimental feasibility of constructing a double-well potential and a cat state for a mirror. Typical experimental parameters are listed in Table \ref{tab: parameters} according to the report in \cite{PhysRevLett.122.071101}. In our feasibility study, we will consider various values of $\kappa$, $\Delta$, $\Omega$, $L$, and $m$. We express them as $(\Omega/2\pi)/(1$ Hz) $=\Omega_0$, $(\Delta/2\pi)/ (10$ kHz$) =\Delta_0$, $(\kappa/2\pi)/(1$ MHz$) =\kappa_0$, $L/(0.06$ m$) =L_0$, and $m/(10$ mg$) =m_0$. For these parameters, the optical frequency shift per unit displacement and the optomechanical coupling are given by $G/2\pi =4.72/L_0$ PHz and $g/2\pi =6.10 g_0$ Hz, where $g_0 =\Omega_0^{-1/2} L_0^{-1}m_0^{-1/2}$. 
For realizing the quantum ground state, we need to increase $\Omega/2\pi \sim 100$ Hz by an additional optical spring effect \cite{PhysRevLett.124.221102}. 

In order to satisfy the condition for the second-order phase transition, we take $\Delta\ll\kappa$. The critical number of photons is given by $n_c\approx 167965 \times \Omega_0^2 \kappa_0^2 L_0^2 {m_0}/\Delta_0$. Any value of $n$ larger than $n_c$ makes the potential double-well shaped. The corresponding input laser power is $P\approx 49.4$ nW $\times \Omega_0^2 \kappa_0^3 L_0^2 {m_0}/\Delta_0$ for $\kappa_i \ll \kappa_e$. The laser wavelength is $2\pi c/\omega_l=1064$ nm, where $c$ is the speed of light.

After preparing the ground state localized at the origin at $n=n_c$, we slowly increase $n$ to construct a cat state. The distance of the cat state $q_{\text{min}}$ is limited by the condition in Eq.\,(\ref{eq: nomixing condition}). Note that $H/(\hbar\Omega)$ depends on two combinations of parameters: $g/\kappa$ and $\Delta/\kappa$. We express the photon number as $n=n_c \lambda$ and numerically calculate the energy gap between the ground state and the second excited state (having the same parity as the ground state), and the overlap of their wavefunctions, as functions of the two parameters. Fig. \ref{fig: mixing adiabatic_mirror}  shows that they have little dependence on $\Delta_0/\kappa_0$. The dependence on $g_0/\kappa_0$ is well-fitted (blue lines in Fig. \ref{fig: mixing adiabatic_mirror}) by power functions with exponents $x \approx 0.672$ and $y \approx 1.36$ as follows: 
\aln{
\Delta E/\hbar &\sim 9.78\times10^{-4}\,(g_0/\kappa_0)^{x}\Omega\nonumber\\
&\sim6.14\times10^{-3}\Omega_0^{1-x/2}(\kappa_0L_0)^{-x}m_0^{-x/2}} 
and 
\aln{|\langle d_{\lambda}\Psi_i|\Psi_0 \rangle |&\sim5.65\times10^5(\kappa_0/g_0)^{y}\nonumber\\
&\sim5.65\times10^5 (\Omega_0 m_0)^{y/2}(\kappa_0L_0)^{y}.
}
Thus the condition Eq.\,(\ref{eq: nomixing condition}) becomes
\aln{
\frac{\dot{n}}{n_c} < 1.09\times10^{-8}\,
\Omega_0^{1-(x+y)/2} \left(\kappa_0 L_0 m_0^{1/2}\right)^{-x-y}.
}
Integrating it over time, the maximally possible value of $\delta n$ which can be changed during the coherence time $t_{\text{coh}}$ is given by
\aln{\delta n =  1.09\times10^{-8}  \Omega_0^{\frac{2-x-y}{2}} \left(\kappa_0 L_0 m_0^{1/2}\right)^{-x-y}
 n_c\, t_{coh}.
\label{eq: delna n}
}
For shorter coherence time $t_{\text{coh}}$, we can vary a smaller amount of $n$, and accordingly, we can construct a cat state with a shorter distance $q_{\text{min}}$ as given by Eq.\,(\ref{eq:nqrelation}). However, $\delta n$ cannot be too small because the number of cavity photons may fluctuate, resulting in an uncertainty $\delta_f n$ due to fluctuations in laser power $F$ or detuning frequency as described in Eq.\,(\ref{eq: determination of n}). Here we assume that the accuracy of the laser power $\delta_f n/n_c$ is given by $a_0 \times 10^{-3}$. The change $\delta n$ must be larger than this uncertainty $\delta_f n$.

Let us now discuss a condition for $\Omega$. 
Large $\Omega$  makes the coherence time shorter. 
At $n=n_c$, it becomes
\aln{
\frac{1}{t_{coh}} \sim \gamma_m +  \frac{\ 4 \Omega^2}{ \kappa} .
}
Then, from Eq.\,(\ref{eq: delna n}), 
the condition $\delta n >\delta_f n$ leads to 
\gat{
\frac{\gamma_m}{2\pi} <  \frac{1.73\times10^{-6}}{a_0\Omega_0^{(x+y)/2-1} (\kappa_0 L_0 m_0^{1/2 })^{x+y}}-\frac{4\times10^{-6}\Omega_0^2}{\kappa_0}.
\label{eq: gamma_bound}
}
The right-hand side of Eq.\,\eqref{eq: gamma_bound} monotonically decreases as $\Omega_0$ increases. Therefore, smaller $\Omega_0$ is preferable, as it allows a larger mechanical friction rate $\gamma_m$. On the other hand, the right-hand side must be positive, imposing an upper bound on the allowed value of $\Omega_0$.
\aln{
\Omega_0<\biggl( \frac{0.433}{ a_0 
 (L_0 m_0^{1/2})^{x+y} \kappa_0^{x+y-1} } \biggr)^{2/(x+y+2)}.
\label{eq: Omega_bound}
}
In experiments, it is preferable to have a large $\Omega_0 \gg 1$ for preparing a ground state wave function.
To achieve a larger value of $\Omega_0$ in Eq.(\ref{eq: Omega_bound}), 
we need a smaller cavity length $L_0$, a lighter mass $m_0$, and a smaller optical loss rate $\kappa_0$.
For instance, if we choose parameters such as $(\kappa_0,L_0, m_0)=(0.1,0.05, 0.1)$, 
the upper bound of $\Omega_0$  is $\Omega_0 = 99.4$ at $a_0=2.0$. 
If we set $\Omega_0 = 80$, the required value of the mechanical friction $\gamma_m/2\pi$ in Eq.(\ref{eq: gamma_bound}) should be smaller than 0.141.  

Then the coherence time $t_{\text{coh}} \sim 1/\gamma_m$ is approximately $1.13$ seconds. During this time, we can create a cat state with a separation of approximately ${2 q_{\text{min}} \lesssim 73 \ a_0^{1/2}}$ by adjusting the laser power such that $\delta n/n_c \sim a_0 \times 10^{-3}$, where $n_c \approx 2687/\Delta_0$. The corresponding Poisson fluctuation is $1/\sqrt{n_c} = 1.93 \times 10^{-2}\Delta_0^{1/2}$. If we can achieve a small detuning such as $\Delta_0 = 0.01$, the fluctuation aligns with our original assumption of $a_0 \times 10^{-3}$ with $a_0 = 2.0$, achievable with a shot-noise limited-laser at 7.9 nW. 

\vspace{1mm}
{\it  Conclusion}.--- 
We proposed an experimental setup for constructing a spatial cat state by introducing dual  cavities with a single shared mirror in the middle. 
Radiation pressure in the left-right symmetric configuration with identical laser parameters generates a double-well shaped potential.  If the parameters satisfy a condition, the deformation of the potential is like a second-order phase transition with respect to laser power. Then, the ground state wave function localized at origin metamorphoses into a spatial cat state in the double-well potential.
 
This estimation relies on two key assumptions. Firstly, we assume that we can initially prepare the ground state wave function localized at the origin. 
Secondly, we assume that parity is preserved in our experimental setup, such that the ground state wave function with even parity does not mix with odd parity states. 
Further investigation into these assumptions is warranted in future studies. Although still presenting challenges, the experimental realization of our proposal seems achievable using state-of-the-art technologies.

{\it Acknowledgements}.--- 
This work is supported in part by Grants-in-Aid for Scientific Research No.18H03708 and JST FORESTO Grant No. JPMJFR202X 
from the Japan Society for the Promotion of Science.


\bibliography{optomech} 

\begin{thebibliography}{10}

\bibitem{Schroedinger1935}
E.~Schr\"{o}dinger.
\newblock Die gegenw\"{a}rtige situation in der quantenmechanik.
\newblock {\em Naturwissenschaften}, 23(48):807--812, Nov 1935.

\bibitem{PhysRevLett.63.612}
Mark~A. Kasevich, Erling Riis, Steven Chu, and Ralph~G. DeVoe.
\newblock rf spectroscopy in an atomic fountain.
\newblock {\em Phys. Rev. Lett.}, 63:612--615, Aug 1989.

\bibitem{tonomura1989}
A.~Tonomura, J.~Endo, T.~Matsuda, T.~Kawasaki, and H.~Ezawa.
\newblock Demonstration of single‐electron buildup of an interference pattern.
\newblock {\em Am. J. Phys.}, 57:117--120, 1989.

\bibitem{Fein2019}
Yaakov~Y. Fein, Philipp Geyer, Patrick Zwick, Filip Kia{\l}ka, Sebastian Pedalino, Marcel Mayor, Stefan Gerlich, and Markus Arndt.
\newblock Quantum superposition of molecules beyond 25 kda.
\newblock {\em Nature Physics}, 15(12):1242--1245, Dec 2019.

\bibitem{friedman2000quantum}
Jonathan~R Friedman, Vijay Patel, Wei Chen, SK~Tolpygo, and James~E Lukens.
\newblock Quantum superposition of distinct macroscopic states.
\newblock {\em nature}, 406(6791):43--46, 2000.

\bibitem{Marshall:2002exi}
William Marshall, Christoph Simon, Roger Penrose, and Dik Bouwmeester.
\newblock {Towards quantum superpositions of a mirror}.
\newblock {\em Phys. Rev. Lett.}, 91:130401, 2003.

\bibitem{PhysRevLett.119.240401}
Sougato Bose, Anupam Mazumdar, Gavin~W. Morley, Hendrik Ulbricht, Marko Toro\ifmmode~\check{s}\else \v{s}\fi{}, Mauro Paternostro, Andrew~A. Geraci, Peter~F. Barker, M.~S. Kim, and Gerard Milburn.
\newblock Spin entanglement witness for quantum gravity.
\newblock {\em Phys. Rev. Lett.}, 119:240401, Dec 2017.

\bibitem{PhysRevLett.119.240402}
C.~Marletto and V.~Vedral.
\newblock Gravitationally induced entanglement between two massive particles is sufficient evidence of quantum effects in gravity.
\newblock {\em Phys. Rev. Lett.}, 119:240402, Dec 2017.

\bibitem{Marquardt_2006}
Florian Marquardt, J.~G.~E. Harris, and S.~M. Girvin.
\newblock Dynamical multistability induced by radiation pressure in high-finesse micromechanical optical cavities.
\newblock {\em Physical Review Letters}, 96(10), March 2006.

\bibitem{PhysRevLett.122.071101}
Nobuyuki Matsumoto, Seth~B. Cata\~no Lopez, Masakazu Sugawara, Seiya Suzuki, Naofumi Abe, Kentaro Komori, Yuta Michimura, Yoichi Aso, and Keiichi Edamatsu.
\newblock Demonstration of displacement sensing of a mg-scale pendulum for mm- and mg-scale gravity measurements.
\newblock {\em Phys. Rev. Lett.}, 122:071101, Feb 2019.

\bibitem{PhysRevLett.124.221102}
Seth~B. Cata\~no Lopez, Jordy~G. Santiago-Condori, Keiichi Edamatsu, and Nobuyuki Matsumoto.
\newblock High-$q$ milligram-scale monolithic pendulum for quantum-limited gravity measurements.
\newblock {\em Phys. Rev. Lett.}, 124:221102, Jun 2020.

\end{thebibliography}
\bibliographystyle{unsrt}

\end{document}